\documentclass[aps,pra,two column,showpacs]{revtex4}
\usepackage{amssymb,amsmath,amsfonts}
\usepackage{bbm,wasysym}
\usepackage{graphicx}
\newcommand{\proj}[1]{{\ket{#1}\bra{#1}}}

\newcommand{\bra}[1]{{\langle #1 |}}
\newcommand{\ket}[1]{{| #1 \rangle}}
\newcommand{\bracket}[2]{{\langle #1 | #2 \rangle}}

\newcommand{\id}{{\mathbbm 1}}

\begin{document}
 \title{Optimal eavesdropping on noisy states
in  quantum key distribution}
 \author{Z. Shadman}
 \email{shadman@thphy.uni-duesseldorf.de}
 \author{H. Kampermann}
 \author{T. Meyer}
 \author{D. Bru\ss}
 \affiliation{Institute f\"ur Theoretische Physik III,
              Heinrich-Heine-Universit\"at
              D\"usseldorf, D-40225 D\"usseldorf, Germany}
 \date{\today}
\begin{abstract}
We study eavesdropping in
quantum key distribution with the six state protocol,
 when the
signal states are mixed with white noise.
This situation may arise either when Alice
deliberately adds noise to the signal states before they leave
her lab, or in a realistic scenario where  Eve cannot
replace the noisy quantum channel by a noiseless one. We
find Eve's optimal mutual information with Alice, for
individual attacks, as a function of the qubit error rate.
Our result is  that  added
quantum noise can make quantum key distribution
  more robust against eavesdropping.
\end{abstract}

\maketitle
%==============================================================================%
\section { Introduction}

In quantum cryptography -- or, more precisely, quantum key distribution --
a secret key is established between two trusted parties (Alice and Bob),
by employing certain quantum states as signals, and suitable measurements
\cite{bb84}. In a typical implementation, polarized photons are sent
from Alice to Bob along an optical fiber \cite{review}.
An eavesdropper (Eve) is usually assumed to have every
possible power that is compatible with the laws of quantum mechanics.
This power is not necessarily realistic, with respect to existing tools
and technology. In particular, Eve is supposed to be able to replace
the quantum channel (i.e. the optical fiber in our example given above),
which in reality always introduces some noise, by a noiseless fiber.
Even though this assumption is compatible with quantum mechanics, it
is too restrictive from a realist's point of view.

In this paper, we will  assume that Eve does not possess
a noiseless fiber. Thus, the quantum states will experience
noise in transit from Alice to Bob. We will consider white
noise, where the noise parameter $p$ describes
the best existing fiber that Eve can possibly get hold of.
Our results
also hold for the scenario where Alice deliberately adds noise to the state
before sending it out of her laboratory, and the quantum channel is noiseless.

Note that we are considering
additional noise at the quantum level. A related question has been
studied in \cite{Renner}: there,  it has been shown that if one
of the parties (Alice or Bob) adds some  noise to their
classical measurement data before  error correction,
 then the BB84 \cite{bb84},
 B92 \cite {B92} and six state protocol \cite{brussQKD,
6states}
are more robust with respect to quantum noise, i.e. the secret
key rate is non-zero up to higher values of the quantum bit error rate.
%One possibility to add noise at the classical level is to add it
%already at quantum level.
Adding noise at the quantum level, i.e. before the measurement, will
also lead to additional noise at the classical level.
In this sense our scenario should lead,
at least qualitatively, to a similar result as shown in \cite{Renner}.

The aim of our paper is to present an intuitive understanding for
the counter-intuitive fact (shown in \cite{Renner})
 that noise may help the trusted parties
to improve the
performance of a quantum cryptographic protocol. We will
derive the optimal mutual information that Eve can obtain, when
using individual attacks on noisy quantum signals, and compare
it to the mutual information achievable by eavesdropping on pure states.
One expects that Alice and Bob, but also Eve, will lose some information, due to the additional noise. It is not evident, however,
how the relation between the two mutual information curves (Alice and Bob versus Alice and Eve ) changes, when the noise increases.

In this paper we will discuss  the
six state protocol with additional
(equal) noise on all signal states. The six signal states of the protocol
with pure states are $\{\ket{0_x},\ket{1_x}, \ket{0_y},\ket{1_y},
\ket{0_z},\ket{1_z}\}$, where $\ket{0_\alpha}$ and $\ket{1_\alpha}$
with $\alpha = x,y,z$ denote
the eigenstates of Pauli operator $\sigma_\alpha$.
 Here, the states $\ket{0_\alpha}$  symbolize the classical bit value 0,
 and $\ket{1_\alpha}$
  represents the classical bit
value 1.

%========================================================================================================================eavesdropping
 \section{Eavesdropping on mixed states}

  In the six state protocol with mixed states
Alice sends instead of pure states  one of the following
six mixed states (either deliberately, or due to unavoidable noise in the
transmission channel):

\begin{eqnarray}
\rho^{\scriptscriptstyle i}={(1-p)}\proj{i}+ \frac{{p}}{2} \id ,
\hspace{3mm} i\in \{ 0_\alpha,1_\alpha \}
% 0,1,\bar{0},\bar{1},\bar{\bar{0}} \hspace{1mm} \&\hspace{1mm} \bar{\bar{1}}.
\end{eqnarray}
with $\alpha = x,y,z$.
The parameter $p$ describes the amount of noise, with $0\leq p\leq 1$.
Here, we assume the noise to be equal in all bases, i.e. 
we study the depolarising channel. (In a more
general model, polarization dependent noise could be
treated in an analogous way, by letting
$p_\alpha$ depend on $\alpha = x,y,z$.)

For the eavesdropping strategy we assume that Eve is restricted
to interfering separately with each of the single systems sent by Alice
(i.e. individual attack). In this  class of attacks she attaches
to each qubit an independent probe which is initially in the state
$\ket{X}$ and applies some unitary transformation.
The dimension of the probes and the interaction are in principle
arbitrary, but in \cite{fush5} it has been shown that
the most general unitary eavesdropping attack on a $d$-dimensional
signal state needs only $d^2$ linearly independent ancilla states
of Eve. (This argument also holds when the signal states are mixed,
as the unitary transformation of the basis states already uniquely
defines the transformation of any superposition, due to linearity,
and thus also of a mixture of projectors onto superpositions of
basis states.) Thus,
it is enough for Eve to use two qubits for her probe states.

 The most general unitary transformation $U$ that Eve can
 design is defined via its action on the basis states
(where we use for the computational basis
the notation $\ket{0}=\ket{0_z}$ and
$\ket{1}=\ket{1_z}$),
 \begin{subequations}
 \begin{eqnarray}
  U\ket{0}\ket{X}=\sqrt{1-D}\ket{0}\ket{A}+\sqrt{D}\ket{1}\ket{B},
\label{unitarya}\\
  U\ket{1}\ket{X}=\sqrt{1-D}\ket{1}\ket{C}+\sqrt{D}\ket{0}\ket{D},
\label{unitaryb}
\end{eqnarray}
\end{subequations}
 where D is called the disturbance, with $0\leq D \leq {\frac{1}{2}}$.
 Eve's normalized probes after
 interaction  are  $\ket{A}$, $\ket{B}$, $\ket{C}$,
and $\ket{D}$.
They have to be chosen such that $U$  is a unitary operator.

The quantum bit error rate in the $z$-basis is denoted as $Q_z$,
and given as the fraction of original signals
 $\ket{0} (\ket{1})$ sent by Alice, but
interpreted as $\ket{1}(\ket{0})$ by Bob, namely
 \begin{eqnarray}
  Q_z=\textstyle \frac{1}{2} \bra 0 \rho_{\scriptscriptstyle
1}^{\scriptscriptstyle B}\ket0 +\frac{1}{2}
\bra1\rho_{\scriptscriptstyle 0}^{\scriptscriptstyle B}\ket1 ,
\end{eqnarray}
where  $ \rho_{\scriptscriptstyle 1}^{\scriptscriptstyle B}$  and
$\rho_{\scriptscriptstyle 0}^{\scriptscriptstyle B} $ are the states
that Bob receives when Alice sends $\ket{0}$ and
$\ket{1}$, respectively. We define  $Q_{x,y}$
in an analogous way for the  $x$ and $y$-basis.

As we assume the noise to be uniform, a quantum bit error rate that
 is basis-dependent indicates the presence of an eavesdropper.
We therefore suppose that Eve uses a strategy that produces
the same quantum bit error rate in the three different
bases, i.e.
\begin{eqnarray}
Q=Q_z=Q_x=Q_y.
\label{equalq}
\end{eqnarray}

 It can be easily verified that
 the relationship between the quantum bit error rate Q and D is
\begin{eqnarray}
Q=D(1-p)+\frac{p}{2}.
\label{qber}
\end{eqnarray}
Additionally, we restrict Eve to attack in such a way
that the two terms of Q are identical, i.e.
 \begin{eqnarray}
\bra0 \rho_{\scriptscriptstyle 1}^{\scriptscriptstyle B}\ket0 =\bra1
\rho_{\scriptscriptstyle 0}^{\scriptscriptstyle B}\ket1  \hspace{1mm},
\label{sym}
 \end{eqnarray}
which can be tested by Alice and Bob, by comparing a part of their
bit string for the $z$-basis.
Again, an analogous requirement has to hold in the other two
bases, too.

Equations (\ref{equalq}) and (\ref{sym}), together with
the unitarity of U lead to the four following
conditions for Eve's states:
\begin{subequations}
\begin{eqnarray}
&\langle B\ket D=0,
\label{7a}\\
\label{constraint1}
& Re\langle A\ket C=\frac{2(1-2Q)}{2-p-2Q}, 
\label{7b}\\
&\langle A\ket B +\langle D\ket C=0, \\
&\langle A\ket D +\langle B\ket C=0.
\label{7d}
\end{eqnarray}
\end{subequations}
Note that the quantum bit error rate $Q$
only depends on the scalar product between $ \ket A$ and $\ket C$.
Eve's two-qubit states can be written as an expansion of four basis vectors
with complex coefficients.
As explained above, Eve's states only need to be four-dimensional.
We have the freedom to choose
$\ket B =\ket {00}$. Equation (\ref{7a}) allows to assign
$\ket D$  one of the other three basis vectors, e.g., $\ket D =\ket {11}$.
The general expansion for  the normalized
vectors $\ket A $ and  $\ket C$ is
\begin{subequations}
\begin{eqnarray}
\ket A =\alpha_{\scriptsize A} \ket {00}+\beta_{\scriptsize A}
\ket{10} +\gamma_{\scriptsize A} \ket{01} +\delta_{\scriptsize A}
\ket{11},
\label{keta}
\\  \text{with}\ \
|\alpha_A|^2+|\beta_A|^2+|\gamma_A|^2+|\delta_A|^2=1 \ ,
\label{constraint2}
\end{eqnarray}
\end{subequations}
and
\begin{subequations}
\begin{eqnarray}
\ket C =\alpha_{\scriptsize C} \ket {00}+\beta_{\scriptsize C}
\ket{10} +\gamma_{\scriptsize C}\ket{01}
+\delta_{\scriptsize C} \ket{11},
\label{ketc} \\
\text{with}\ \
|\alpha_C|^2+|\beta_C|^2+|\gamma_C|^2+|\delta_C|^2=1 .
\label{constraint3}
\end{eqnarray}
\end{subequations}
We have to determine the free parameters $\alpha_A,...,\delta_A$
and $\alpha_C,...,\delta_C$ such that Eve's transformation is
optimized. As a figure of merit we will calculate
the  mutual information
between Eve and Alice, and optimize Eve's transformation such that
she acquires the maximal mutual information.

%============================================================================================================ RESULTS
\section { RESULTS}

The mutual information measures the information that two parties share.
Here the parties have variables
$X,Y$ that can take values $x,y$, respectively. The mutual information
 is defined \cite {Nielsen} as
\begin{eqnarray}
 I^{XY} =\sum_{x,y}p(x,y)\log p(y|x)-\sum_{y}p(y)\log p(y) ,
\label{defmutualinfo}
\end{eqnarray}
where $p(x,y)$ is the joint probability to find
$x$ and $y$, and $p(y|x)$ is the conditional probability of $y$, given $x$.
All logarithms are taken to base 2.

 Eve wishes to
retrieve  the maximal information, i.e. she has to choose the
optimal coefficients
$\alpha_{A,C},\beta_{A,C},\gamma_{A,C},\delta_{A,C}$,
 for fixed $p$ and $Q$.
The full problem can be simplified with the following argument:
As  mentioned above,
for a fixed noise parameter $p$ the  quantum bit
error rate $Q$ only depends on the real part of the
overlap between  $\ket A $ and $\ket C$, see equation (\ref{7b}).
Therefore, Eve is free to choose those states on which $Q$ does
not depend in such a way that her information is maximal,
as long as the constraints given in equations (\ref{7a})-(\ref{7d})
are fulfilled. Thus, Eve will choose  her ancilla
states orthogonal (whenever possible), i.e.
$\bracket{A}{B}=\bracket{B}{C}=\bracket{A}{D}=\bracket{D}{C}=0$,
which corresponds to
$\alpha_A =\delta_A=\alpha_C=\delta_C=0$.
One realizes by looking at equation (\ref{unitarya}) and (\ref{unitaryb})
that in this way Eve's probe states are made as
distinguishable as possible (for given $\bracket{A}{C}\neq 0$).
The best measurement for the two remaining non-orthogonal states
$\ket{A}$ and $\ket{C}$ is rank one and orthogonal
\cite{davies,Nielsen}.

With the above definition (\ref{defmutualinfo}) for the
mutual information  besides the explicit
expansions of Eve's states as in (\ref{keta}), (\ref{ketc}),
the information that Eve  acquires is
\begin{eqnarray}
I^{AE}&=&1+{\frac{1}{2}}(\frac{1-\frac{p}{2}-Q}{1-p}) \cdot \nonumber\\
&\cdot &\left\{
\tau\Big[(1-{\frac{p}{2}})|\beta_A|^2+{\frac{p}{2}}|\beta_C|^2
\right.
{\bf,}{\frac{p}{2}}|\beta_A|^2+(1-{\frac{p}{2}})|\beta_C|^2\Big]\nonumber\\
&+&
%{\frac{1}{2}}(\frac{1-\frac{p}{2}-Q}{1-p})
\tau\Big[(1-{\frac{p}{2}})|\gamma_A|^2+{\frac{p}{2}}|\gamma_C|^2
{\bf,}\left. {\frac{p}{2}}|\gamma_A|^2+(1-{\frac{p}{2}})|\gamma_C|^2\Big]
 \right\}\nonumber\\
&+& (\frac{Q-\frac{p}{2}}{1-p}).\tau \Big[
1-{\frac{p}{2}}
\;\; {\bf ,} \;\;
 {\frac{p}{2}}
\Big],
\end{eqnarray}
where we used the definition
\begin{eqnarray}
\tau[x,y]=x \log x+ y\log y -(x+y)\log(x+y).
\end{eqnarray}
  We used the method of Lagrange multipliers
for the optimization problem. Some details and the
equations to be solved are given in the Appendix.

As sketched in the Appendix, we arrive at
$ |\beta_C|^2= 1-|\beta_A|^2$. Thus, we find that the maximal mutual
information between Alice \& Eve is
\begin{eqnarray}
 I^{AE}&=&1+(\frac{1-\frac{p}{2}-Q}{1-p}).\nonumber\\
&.&\Big\{ \big({(1-p)}|\beta_A|^2+{\frac{p}{2}}\big)\log\big({(1-p)}|\beta_A|^2
+{\frac{p}{2}}\big) \nonumber\\
&+&\big({1-\frac{p}{2}}-{(1-p)}|\beta_A|^2\big)\log\big({1-\frac{p}{2}}-{(1-p)}|\beta_A|^2\big)\Big\}\nonumber\\
&+&(\frac{Q-\frac{p}{2}}{1-p})\Big\{{\frac{p}{2}}\log {\frac{p}{2}}+(1-{\frac{p}{2}})\log {(1-\frac{p}{2})}\Big\},
\label{solution}
\end{eqnarray}
where
\begin{eqnarray}
 &{|\beta_A|^2 ={
 \frac{1}{2}}\Big(1+\frac{1}{1-\frac{p}{2}-Q}\sqrt {(Q-\frac{p}{2})(2-3Q-\frac{p}{2})}\;\Big)}.\label{14}
\end{eqnarray}
Obtaining the mutual information between Alice and Bob is an easy task.
From the definition of the mutual information we find
\begin{eqnarray}
I^{AB}= 1 + Q \log Q+ (1-Q) \log (1-Q).
\end{eqnarray}
One easily confirms that in the absence of white noise, i.e. for $p=0$,
 the mutual information functions
reduce to the noiseless case described in \cite{brussQKD}.

We have plotted the mutual information curves
$I^{AB}$ and $I^{AE}$ as a function of the qubit error rate $Q$
 in Fig. 1, for an example with $p=0.05$.
Note that for $p\neq 0$ both information curves start at a non-zero
value for $Q$, as we have $Q\geq p/2$ from equation (\ref{qber}).
 In comparison to the noiseless case
the mutual information between Alice \& Eve is lower
than without noise. As $p$ increases, $I^{AE}$ decreases.
On the other hand, for increasing $p$, the mutual information curve
for Alice \& Bob starts at higher values for $Q$, but otherwise
remains invariant. Intuitively speaking, both the trusted and untrusted
parties undergo some degradation of their information, due to
the noise. Which consequence does this have for the creation of a secret
key?

Let us turn our attention to the crossing point between the two mutual
information curves.
A secret key can be established if $ I^{AB}-I^{AE}\ge 0$ \cite{csisza`r},
i.e. for values of $Q$ which are smaller than the value for the
crossing point. In Fig. 1 we observed that for  one example of non-zero
$p$ the crossing-point moves towards a larger $Q$. We therefore
studied the $Q$-value for the crossing-point as a function of $p$.
The result is shown in
 Fig. 2. Remember the relation between  qubit
error rate $Q$, disturbance $D$ and noise $p$, given in equation (\ref{qber}). 
One might expect that
 the crossing point of the two information curves
obeys this  linear dependence, i.e. $Q_{cross}= D_{cross}(1-p)+p/2$; this
is  the dashed line in   Fig. 2. However, the true value
for the crossing point lies
above that straight line. Thus,
as $p$ increases, the crossing point moves to a higher
quantum bit error rate than expected from the additional noise. 
So, by adding  noise to the quantum data
the secure parameter range of $Q$ will increase.
 In other words, the six state protocol with mixed states is
more robust against eavesdropping  than the six  state protocol with pure
states.

In summary, we have studied individual eavesdropping attacks on
the six state protocol with added white noise. We found the optimal
mutual information between Eve and Alice, for a fixed amount of noise,
as a function of the qubit error rate. We showed that the crossing
point between the mutual information curves for Alice/Bob and
Alice/Eve  (Csisz\'ar-K\"orner threshold)
moves to a qubit error rate that is higher than expected from
the noise alone.
Thus, we gave a simple explanation for the counter-intuitive
fact that added noise may
improve the robustness of a quantum key distribution protocol.
Our results are of importance in a realistic scenario, where the
eavesdropper does not have a noiseless fiber available.
Alternatively, Alice can deliberately add noise to
the signal states.
In this paper, we have focused on a particular QKD protocol and
a particular type of noise.
As an
outlook, it would be interesting to study other protocols, and/or
other types of noise.

{\em Acknowledgements:} We acknowledge stimulating discussions
with Matthias Kleinmann and Chiara Macchiavello.
 This work was
partially supported by the EU Integrated Projects SECOQC and SCALA.

\newpage

%============================================================================================================================================BIBLIOGRAPHY
\begin{figure}[h]
\includegraphics[width=7.5cm]{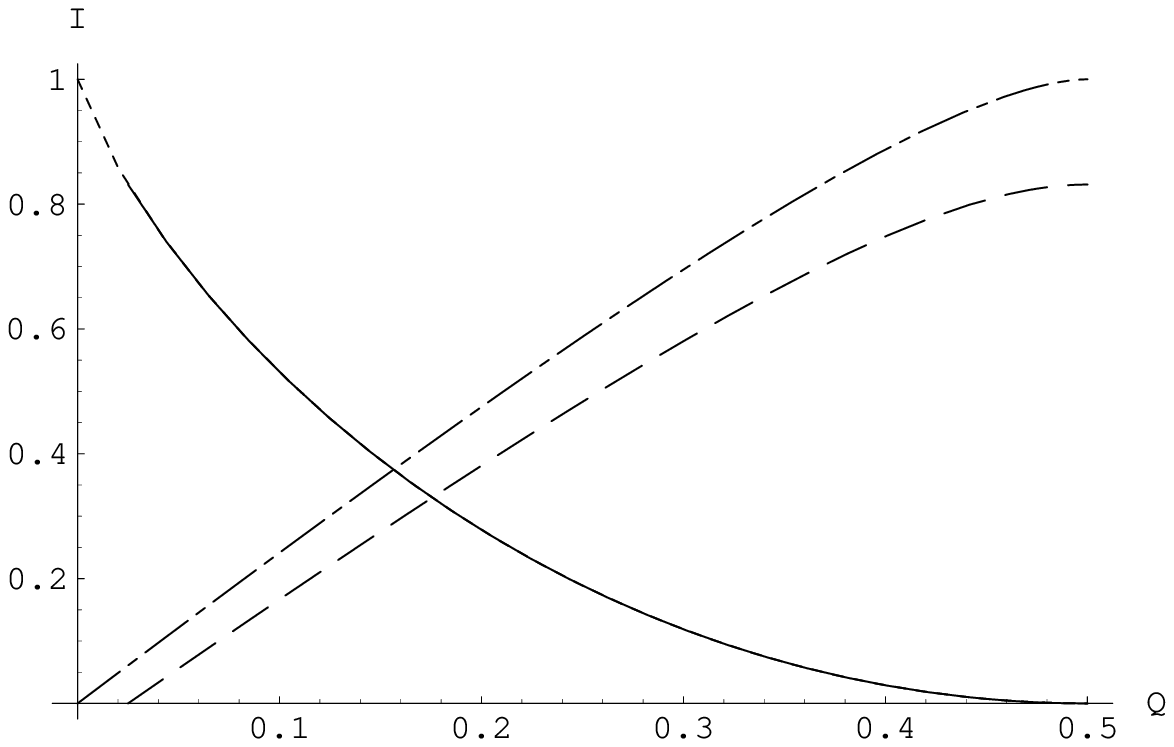}

\_ \_ \_   {\scriptsize   $I^{AB}$ for six pure states}

------  {\scriptsize $I^{AB}$ for six mixed states}

--- - ---   {\scriptsize Optimal $I^{AE}$  for six pure states}

$-$     $-$      $-$  {\scriptsize Optimal $I^{AE} $ for six mixed states}
\end{figure}
 FIG. 1.{ \scriptsize
Mutual information between Alice \& Bob and Alice \& Eve for six pure and six mixed state cases, as a function of qubit error rate (Q) when noise parameter is $p=0.05$}.

\vspace*{1cm}
\begin{figure}[htb]
\includegraphics[width=9cm]{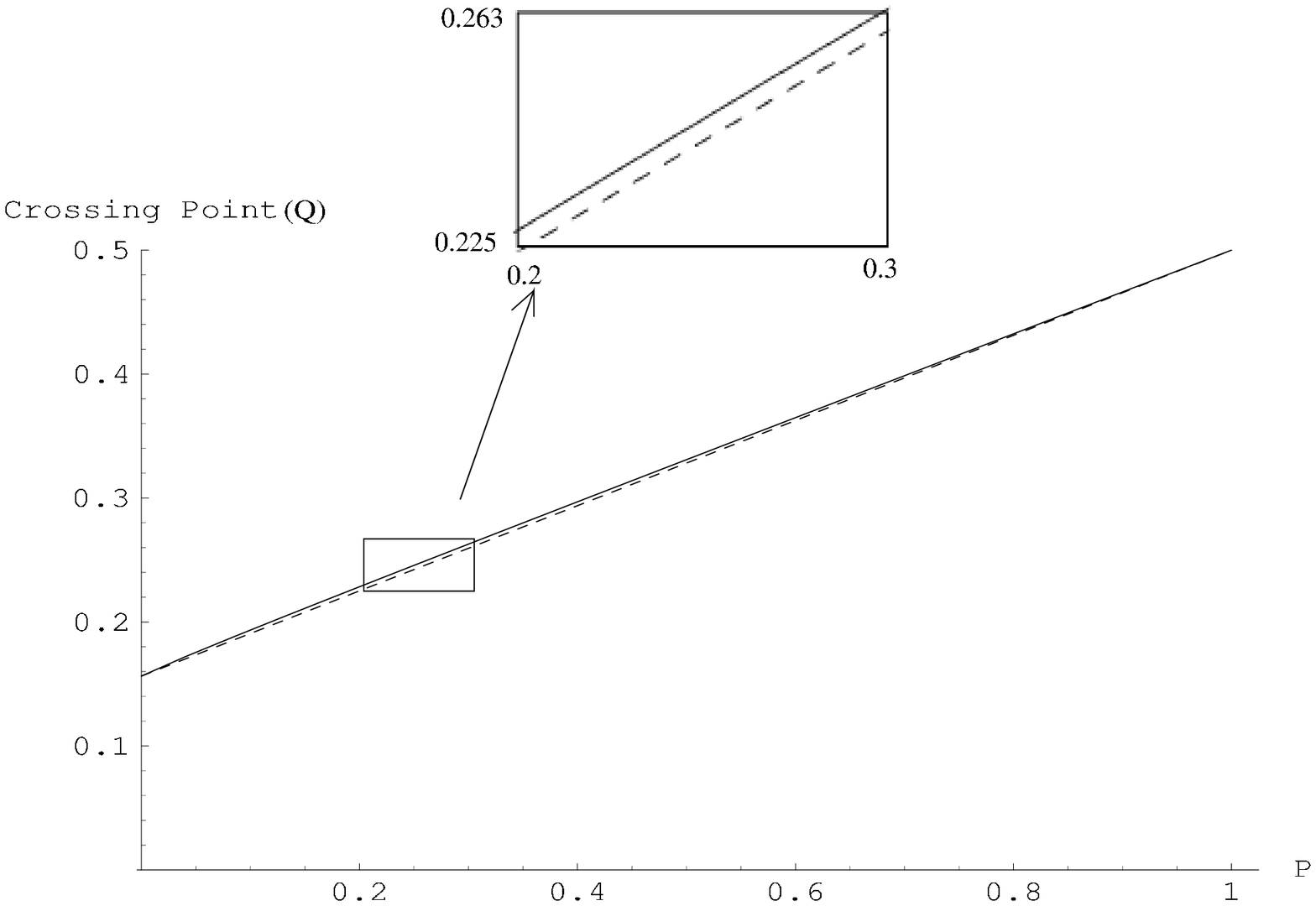}
\end{figure}
FIG. 2. { \scriptsize The solid line is the value of $Q$ for the crossing point of $I^{AB}$ and $I^{AE}$, as a function of the 
noise parameter $p$.
 The dashed line is the straight line that corresponds to 
equation (\ref{qber}) when D=0.15637.}
%==========================================================================================================================================
%\clearpage
\vspace*{1cm}
\section{Appendix}
Here we briefly explain our method for obtaining the optimized $I^{AE}$.
First, we redefine the complex coefficients
$\beta_{A,C},\gamma_{A,C}$
in polar coordinates:
\begin{eqnarray}
\beta_A&=&{\Large r}_{\beta_A},\\
\beta_C&=&{\Large r}_{\beta_C},\\
\gamma_A&=&{\Large r}_{\gamma_A}\exp (i\Phi_{\gamma_A}) ,\\
\gamma_C&=&{\Large r}_{\gamma_C}\exp (i\Phi_{\gamma_C}) .
\end{eqnarray}
 Note that because of the unphysical  global phase we have the
freedom to choose $\beta_A $ and $\beta_C $  real.
Using the Lagrange multiplier method we then write the Lagrangian L as
 \begin{eqnarray}
L=I^{AE}+\lambda_1 g_1+\lambda_2 g_2+\lambda_3 g_3,
\end{eqnarray}
where $g_1 $,$g_2 $ and $g_3$ are the constraints (\ref{constraint1}),(\ref{constraint2}) and (\ref{constraint3}).
%{\tt which ones? refer to equations in text!}%
The derivative $dL = 0$ yields the following system of equations:
\begin{eqnarray}
&&g_1=r_{\beta_A}r_{\beta_C}
+r_{\gamma_A } r_{\gamma_C}\cos (\Phi_{\gamma_A}-\Phi_{\gamma_C})
\nonumber \\
&&\ \ \ \ -\frac{2(1-2Q)}{2-p-2Q}=0,\label{21}\\
&&g_2={r_{\beta_A}}^2+{r_{\gamma_A }}^2-1=0,\label{22}\\
&&g_3={r_{\beta_C}}^2+{r_{\gamma_C}}^2-1=0, \label{23}\\
&&r_{\gamma_A } r_{\gamma_C} \sin (\Phi_{\gamma_A}-\Phi_{\gamma_C})= 0,\label{24} \\
&&r_{\beta_A } \{(\frac{1-\frac{p}{2}-Q}{1-p})((1-\frac{p}{2})\log M_2+ \frac{p}{2}\log M_6\nonumber \\
&&-\log (M_2+M_6))+2\lambda_2\}+\lambda _1 r_{\beta_C}=0,\label{25}\\
&&r_{\beta_C } \{(\frac{1-\frac{p}{2}-Q}{1-p})(\frac{p}{2}\log M_2+ (1-\frac{p}{2}) \log M_6\nonumber \\
&&-\log (M_2+M_6))+2\lambda_3\}+\lambda _1r_{\beta_A}=0,\label{26}\\
&&r_{\gamma_A } \{(\frac{1-\frac{p}{2}-Q}{1-p})((1-\frac{p}{2})\log M_3 \nonumber \\
&&+\frac{p}{2} \log M_7-\log (M_3+M_7))+2\lambda_2\}\nonumber \\
&&+\lambda _1 r_{\gamma_C } \cos (\Phi_{\gamma_A }-\Phi_{\gamma_C})=0,\label{27}\\
&&r_{\gamma_C} \{(\frac{1-\frac{p}{2}-Q}{1-p})(\frac{p}{2}\log M_3 \nonumber \\
&&+(1-\frac{p}{2})\log M_7-\log (M_3+M_7))+2\lambda_3\}\nonumber\\
&&+\lambda _1r_{\gamma_A }\cos (\Phi_{\gamma_A }-\Phi_{\gamma_C})=0.\label{28}
\end{eqnarray}
 Here, $M_1,...,M_4 $  and  $ M_5,...,M_8 $ are the probabilities
that Eve detects $\ket {00},\ket {10},\ket {01}$ and $\ket {11} $ while Alice sends $\rho_{\scriptscriptstyle 0}$ and
$\rho_{\scriptscriptstyle 1}$, respectively. The $M_i$  are defined as
\begin{eqnarray}
M_1=M_8&=&(1-\frac{p}{2}) \cdot
\frac{Q-\frac{p}{2}}{1-p}, \nonumber  \\
M_2&=&(\frac{1-\frac{p}{2}-Q}{1-p})\{(1-\frac{p}{2}){r_{\beta_A}}^2+\frac{p}{2} {r_{\beta_C}}^2\},   \nonumber \\
M_3&=&(\frac{1-\frac{p}{2}-Q}{1-p})\{(1-\frac{p}{2}){r_{\gamma_A }}^2+\frac{p}{2} {r_{\gamma_C}}^2\},  \nonumber\\
M_4=M_5&=&
\frac{p}{2}\cdot \frac{Q-\frac{p}{2}}{1-p}, \nonumber \\
M_6&=&(\frac{1-\frac{p}{2}-Q}{1-p})\{(1-\frac{p}{2}){r_{\beta_C}}^2+\frac{p}{2} {r_{\beta_A}}^2\},  \nonumber\\
M_7&=&(\frac{1-\frac{p}{2}-Q}{1-p})\{(1-\frac{p}{2}){r_{\gamma_C}}^2+\frac{p}{2} {r_{\gamma_A }}^2\}. \nonumber\\
\end{eqnarray}
 It is not straightforward to extract the solution from this
 set of equations.
We will follow a strategy based on analytical and numerical methods. 
Due to equation (\ref{24}) there are two possible solutions,
 which are $\cos (\Phi_{\gamma_A }-\Phi_{\gamma_C})=1$ and
$\cos (\Phi_{\gamma_A }-\Phi_{\gamma_C})=-1$ (as 
$ r_{\gamma_A }$ and $ r_{\gamma_C}$ cannot be zero).

Let us first assume the option $\cos (\Phi_{\gamma_A }-\Phi_{\gamma_C})=1$.
Note that in this case the set of equations (21) - (28) is invariant under
the simultaneous exchange $r_{\beta_A} \leftrightarrow r_{\gamma_A}$
and $r_{\beta_C} \leftrightarrow r_{\gamma_C}$. 
 As we can see from equation (11), the mutual information function 
is also symmetric under this exchange.
Now, 
 we combine the set of the  equations (\ref{21})-(\ref{28})
 to one joint equation in terms of $p$, $Q$ and ${r_{\beta_A}}^2$.
The task is to find all roots for ${r_{\beta_A}}^2$.
From equations (22) and (23) we have ${r_{\gamma_A }}^2=1-{r_{\beta_A}}^2$ 
and ${r_{\gamma_C }}^2=1-{r_{\beta_C}}^2$. This fact, together with
the symmetry mentioned above, means that there has to be an even number
of roots for ${r_{\beta_A}}^2$ (if one finds a solution  for $r_{\beta_A}^2$,
then $1-r_{\beta_A}^2$ is also a solution).  
 Numerically (by plotting the joint equation
in terms of ${r_{\beta_A}}^2$) we show
 that for different $p$ and  $Q$ there are  always exactly two roots.
%Solving equation (\ref{21}) in terms of $r_{\beta_C}$ gives us two following equations for $r_{\beta_C}$,
%\begin{eqnarray}
%\label{sym1}
%& r_{\beta_C}= a r_{\beta_A}+\sqrt{a^2{r_{\beta_A}}^2-a^2-{r_{\beta_A}}^2+1},  \\
%\label{sym2}
%& r_{\beta_C}= a r_{\beta_A}-\sqrt{a^2{r_{\beta_A}}^2-a^2-{r_{\beta_A}}^2+1}.
%\end{eqnarray}
%where $a=\frac{2(1-2Q)}{2-p-2Q}$.
%Since the equation (\ref{21}) is a symmetric equation both of these 
%solutions are acceptable.  
Analytically, ${r_{\beta_C}}^2=1-{r_{\beta_A}}^2$ is a possible 
solution for the equations (\ref{22})-(\ref{28}). By inserting 
this expression for
${r_{\beta_C}}^2$ as well as ${r_{\gamma_A}}^2$ 
and ${r_{\gamma_C}}^2$ (see above) 
into (\ref{21}) we find two solutions for 
${r_{\beta_A}}^2$ which are parametrized in terms of $p$ and $Q$. 
One of them is the equation (\ref{14}), already given in the text,
 and the other one is
\begin{eqnarray}
&{r_{\beta_A}}^2={|\beta_A|^2 ={
 \frac{1}{2}}\Big(1-\frac{1}{1-\frac{p}{2}-Q}\sqrt {(Q-\frac{p}{2})(2-3Q-\frac{p}{2})}\;\Big)}.\nonumber  \\
\end{eqnarray}
Both of them lead to the same mutual information (this is clear from the
symmetry, as explained above). 
Hence, we arbitrarily chose the one given in equation (\ref{14}). 
Comparing analytical and numerical results made us sure that 
${r_{\beta_C}}^2=1-{r_{\beta_A}}^2$ is 
the unique relation between ${r_{\beta_A}}^2$ and  ${r_{\beta_C}}^2$. 

For the case $\cos (\Phi_{\gamma_A }-\Phi_{\gamma_C})=-1$, 
we repeat the above process.
However, equation (\ref{21}) is now {\em not} symmetric under the
exchange $r_{\beta_A} \leftrightarrow r_{\gamma_A}$
and $r_{\beta_C} \leftrightarrow r_{\gamma_C}$. 
If we plot the joint function for equations (\ref{21})-(\ref{28}) 
in terms of ${r_{\beta_A}}^2$, we just find one root,
and thus just expect one solution of the set of equations.
Analytically we obtain ${r_{\beta_C}}^2={r_{\beta_A}}^2$ as a 
possible solution. This leads to the 
following mutual information between Alice and Eve:
\begin{eqnarray}
\label{muinfo(-1)}
I^{AE}=(\frac{Q-\frac{p}{2}}{1-p})\Big\{1+{\frac{p}{2}}\log {\frac{p}{2}}+(1-{\frac{p}{2}})\log {(1-\frac{p}{2})}\Big\}.
\end{eqnarray}
Comparing   the equations
(\ref{solution})  and (\ref{muinfo(-1)}) 
analytically, we see that 
 for all $p$ and $Q$  the function in (\ref{solution}) is bigger
than (\ref{muinfo(-1)}). Therefore,
 equation (\ref{solution}) is the optimal mutual information.
%================================================================= REFERENCE

\end{document}